# Machine Learning for Networking: Workflow, Advances and Opportunities

Mowei Wang, Yong Cui, Xin Wang, Shihan Xiao, and Junchen Jiang

## Abstract

Recently, machine learning has been used in every possible field to leverage its amazing power. For a long time, the networking and distributed computing system is the key infrastructure to provide efficient computational resources for machine learning. Networking itself can also benefit from this promising technology. This article focuses on the application of MLN, which can not only help solve the intractable old network questions but also stimulate new network applications. In this article, we summarize the basic workflow to explain how to apply machine learning technology in the networking domain. Then we provide a selective survey of the latest representative advances with explanations of their design principles and benefits. These advances are divided into several network design objectives and the detailed information of how they perform in each step of MLN workflow is presented. Finally, we shed light on the new opportunities in networking design and community building of this new inter-discipline. Our goal is to provide a broad research guideline on networking with machine learning to help motivate researchers to develop innovative algorithms, standards and frameworks.

## Introduction

With the prosperous development of the Internet, networking research has attracted a lot of attention in the past several decades both in academia and industry. Researchers and network operators can face various types of networks (e.g., wired or wireless) and applications (e.g., network security and live streaming [1]). Each network application also has its own features and performance requirements, which may change dynamically with time and space. Because of the diversity and complexity of networks, specific algorithms are often built for different network scenarios based on the network characteristics and user demands. Developing efficient algorithms and systems to deal with complex problems in different network scenarios is a challenging task.

Recently, machine learning (ML) techniques have made breakthroughs in a variety of application areas, such as bioinformatics, speech recognition and computer vision. Machine learning tries to construct algorithms and models that can learn to make decisions directly from data without following pre-defined rules. Existing machine learning algorithms generally fall into three categories: supervised learning (SL), unsupervised learning (USL) and reinforcement learning (RL). More specifically, SL algorithms learn to conduct classification or regression tasks from labeled data, while USL algorithms focus on classifying the sample sets into different groups (i.e., clusters) with unlabeled data. In RL algorithms, agents learn to find the best action series to maximize the cumulated reward (i.e., objective function) by interacting with the environment. The latest breakthroughs, including deep learning (DL), transfer learning and generative adversarial networks (GAN), also provide potential research and application directions in an unimaginable fashion.

Dealing with complex problems is one of the most important advantages of machine learning. For some tasks requiring classification, regression and decision making, machine learning may perform close to or even better than human beings. Some examples are facial recognition and game artificial intelligence. Since the network field often sees complex problems that demand efficient solutions, it is promising to bring machine learning algorithms into the network domain to leverage the powerful ML abilities for higher network performance. The incorporation of machine learning into network design and management also provides the possibility of generating new network applications. Actually, ML techniques have been used in the network field for a long time. However, existing studies are limited to the use of traditional ML attributes, such as prediction and classification. The recent development of infrastructures (e.g., computational devices like GPU and TPU, ML libraries like Tensorflow and Scikit-Learn) and distributed data processing frameworks (e.g., Hadoop and Spark) provides a good opportunity to unleash the magic power of machine learning for pursuing the new potential in network systems.

Specifically, machine learning for networking (MLN) is suitable and efficient for the following reasons. First, as the best known capabilities of ML, classification and prediction play basic but important roles in network problems such as intrusion detection and performance prediction [1]. In addition, machine learning can also help decision making, which will facilitate network scheduling [2] and parameter adaptation [3, 4], according to the current states of the environment. Second, many network problems need to interact with complicated system environments. It is not easy to build accurate or analytic models to represent complex system behaviors such as load changing patterns of CDN [5] and throughput characteristics [1]. Machine learning can provide an estimated model of these systems with acceptable accuracy. Finally, each network scenario may have different characteristic (e.g., traffic patterns and network states) and researchers often need to

Mowei Wang, Yong Cui, and Shihan Xiao are with Tsinghua University. Yong Cui is the corresponding author.

Xin Wang is with Stony Brook University.

Junchen Jiang is with Carnegie Mellon University.

      

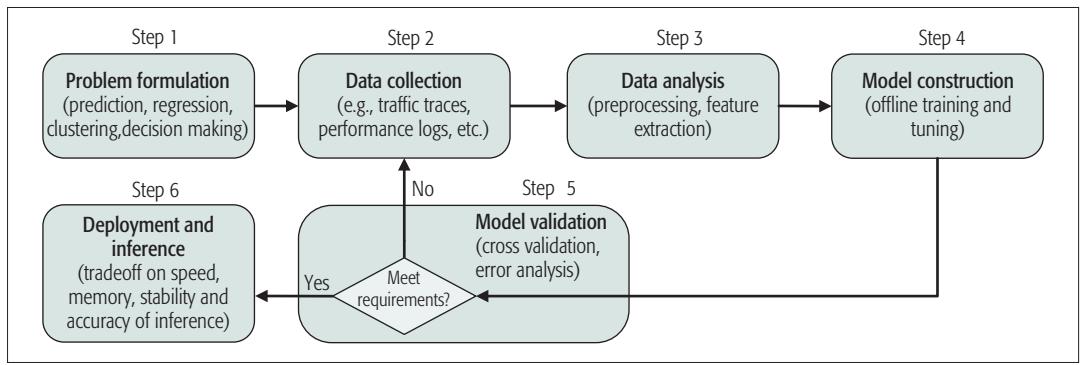

FIGURE 1. The typical workflow of machine learning for networking.

solve the problem for each scenario independently. Machine learning may provide new possibilities to construct the generalized model via a uniform training method [3, 4]. Among efforts in MLN, deep learning has also been investigated and applied to provide end-to-end solutions. The latest work in [6] conducts a comprehensive survey on previous efforts that apply deep learning technology in network related areas.

In this article, we investigate how machine learning technology can benefit network design and optimization. Specifically, we summarize the typical workflows and requirements for applying machine learning techniques in the network domain, which could provide a basic but practical guideline for researchers to have a quick start in the area of MLN. Then we provide a selective survey of the important networking advances with the support of machine learning technology, most of which have been published in the last three years. We group these advances into several typical networking fields and explain how these prior efforts perform at each step of the MLN workflow. Then we discuss the opportunities of this emerging inter-discipline area. We hope our studies can serve as a guide for potential future research directions.

## Basic Workflow for MLN

Figure 1 shows the baseline workflow for applying machine learning in the network field, including problem formulation, data collection, data analysis, model construction, model validation, deployment and inference. These stages are not independent but have inner relationships. This workflow is very similar to the traditional workflow for machine learning, as network problems are still applications that machine learning can play a role in. In this section, we explain each step of the MLN workflow with representative cases.

**Problem Formulation:** Since the training process of machine learning is often time consuming and involves high cost, it is important to correctly abstract and formulate the problem at the first step of MLN. A target problem can be classified into one of the machine learning categories, such as classification, clustering and decision making. This helps decide what kind of and the amount of data to collect and the learning model to select. An improper problem abstraction may provide an unsuitable learning model, which can result in unsatisfactory learning performance. For example, it is better to cast the optimal quality of experience (QoE) for live streaming into a real-time exploration-exploitation process rather than as a prediction-based problem [7] to well match the application characteristics.

**Data Collection:** The goal of this step is to collect a large amount of representative network data without bias. The network data (e.g., traffic traces and session logs with performance metrics) are recorded from different network layers according to the application needs. For example, the traffic classification problem often requires datasets containing packet-level traces labeled with corresponding application classes [8]. In the context of MLN, data are often collected in two phases. In the offline phase, collecting enough high-quality historical data is important for data analysis and model training. In the online phase, real-time network state and performance information are often used as inputs or feedback signals for the learning model. The newly collected data can also be stored to update the historical data pool for model adaption.

**Data Analysis:** Every network problem has its own characteristics and is impacted by many factors, but only several factors (i.e., feature) have the most effect on the target network performance metric. For instance, RTT and the inter-arrival time of ACK may be the critical features in choosing the best size of the TCP congestion window [3]. In the learning paradigm, finding proper features is the key to fully unleashing the potential of data. This step attempts to extract the effective features of a network problem by analyzing the historical data samples, which can be regarded as a feature engineering process in the machine learning community. Before feature extraction, it is important to preprocess and clean raw data, through processes such as normalization, discretization, and missing value completion. Extracting features from cleaned data often needs domain-specific knowledge and insights of the target network problem [5], which is not only difficult but time-consuming. Thus in some cases deep learning can be a good choice to help automate feature extractions [2, 6].

**Model Construction:** Model construction involves model selection, training and tuning. A suitable learning model or algorithm needs to be

> These stages are not independent but have inner relationships. This workflow is very similar to the traditional workflow for machine learning, as network problems are still applications that machine learning can play a role in.

selected according to the size of the dataset, typical characteristics of a network scenario, the problem category, and so on. For example, accurate throughput prediction can improve the bitrate adaption of Internet video, and a Hidden-Markov model may be selected for prediction due to the dynamic patterns of stateful throughput [1]. Then the historical data will be used to train a model with hyper-parameter tuning, which will take a long period of time in the offline phase. The parameter tuning process still lacks enough theoretical guidance, and often involves a search in a large space to find acceptable parameters or to tune by personal experiences.

**Model Validation:** Offline validation is an indispensable step in the MLN workflow to evaluate whether the learning algorithm works well enough. During this step, cross validation is usually used to test the overall accuracy of the model in order to show if the model is overfitting or under-fitting. This provides good guidance on how to optimize the model, e.g., increasing the data volume and reducing model complexity when there exists overfitting. Analyzing wrong samples helps find the reasons for errors to determine whether the model and the features are proper or the data are representative enough for a problem [5, 8]. The procedures in the previous steps may need to be re-taken based on the error sources.

**Deployment and Inference:** When implementing the learning model in an operational network environment, some practical issues should be considered. Since there are often limitations on computation or energy resources and requirements on the response time, the tradeoff between accuracy and the overhead is important for the performance of the practical network system [7]. In addition, machine learning often works in a best-effort way and does not provide any performance guarantee, which requires system designers to consider fault tolerance. Finally, practical applications often require the learning system to take real-time input, and obtain the inference and output the corresponding policy online.

| Networking application | | | Steps of MLN workflow | | | | |
|---|---|---|---|---|---|---|---|
| | | | Data collection | | | | |
| Objectives | Specific works | Problem formulation | Offline collection | Online measurement | Data analysis | Offline model construction | Deployment and online inference |
| Information cognition | Sibyl [11]: route measurement | SL: prediction with RuleFit | Combine data of platforms with a few powerful VPs in homogeneous deployment and with many limited VPs around the world | Take users' queries as input round by round | / | Construct RuleFit model to assign confidence to each predicted path | Optimize measurement budget in each round to get the best query coverage |
| Traffic prediction | Ref [9]: traffic volume prediction | SL: prediction with Hidden-Markov Model (HMM) | Synthetic and real traffic traces with flow statistics | Only observe the flow statistics | The flow count and the traffic volume have significant correlation | Training HMM model with Kernel Bayes Rule and Recurrent Neural Network with Long Short Term Memory unit | Take flow statistics as input and obtain the output of the traffic volume |
| Traffic classification | RTC [8]: traffic classification | SL and USL: clustering and classification | Labeled and unlabeled traffic traces | Flow statistical features extracted from traffic flows | Zero-day-application exists and may degrade the classification accuracy | Find the Zero-day-application class and training the classifier | Inference with the trained model to output the classification results |
| Resource management | DeepRM [13]: job scheduling | RL: decision making with deep RL | Synthetic workload with different patterns is used for training | The real time resource demand of the arrival job | Action space is too large and may has conflicts between actions | Offline training to update the policy network | Directly schedule the arrival jobs with the trained model |
| Network adaption | Ref [2]: routing strategy | SL: decision making with Deep Belief Architectures (DBA) | Traffic patterns labeling with routing paths computed by OSPF protocol | Online traffic patterns in each router | It is difficult to characterize the input and output patterns to reflect the dynamic nature of large-scale heterogeneous networks | Take the Layer-Wise training to initialize and the backpropagation process to fine-tune the DBA structure | Record and collect the traffic patterns in each router periodically and obtain the next routing nodes from the DBAs |
| | Pytheas [7]: general QoE optimization | RL: decision making with a variant of UCB algorithm | Session quality information with features in large time scale | Session quality information in small time scale | Application sessions sharing the same features can be grouped | Backend cluster determines the session groups using CFA [5] with a long time scale | Frontend performs the group-based exploration-exploitation strategy in real time |
| | Remy [3]: TCP congestion control | RL: decision making with a tabular method | Collect experience from network simulator | Calculate network state variables with ACK | Select the most influential metrics as state variables | Given network assumption the generated algorithm interact with simulator to learn best actions according to states | Directly implement the Remy-generated algorithm to corresponding network environment |
| | PCC [4]: TCP congestion control | RL: decision making with online learning | / | Calculate the utility function according the received SACK | TCP assumptions are often violated. The direct performance is a better signal | / | Take trials with different sending rates and find the best rate according to the feedback utility function |
| Performance prediction | CFA [5]: video QoE optimization | USL: clustering with self-designed algorithm | Datasets consisting of quality measurements are collected from public CDNs | Take session features as input, such as Bitrate, CDN, Player, etc. | Similar sessions are with similar quality determined by critical features | Critical feature learning in minutes scale and quality estimation in tens of seconds | Look up feature-quality table to respond to real-time query |
| | CS2P [1]: throughput prediction | SL: prediction with HMM | Datasets of HTTP throughput measurement from iQIYI | Take users's session features as input | Sessions with similar features tend to behave in related pattern | Find set of critical feature and learn a HMM for each cluster of similar sessions | A new session is mapped to the most similar session cluster and corresponding HMM are used to predict throughput |
| Configuration extrapolation | cherryPick [15]: cloud configurations extrapolation | SL: parameter searching with Bayesian optimization | Take performance under current configuration as model input | / | Large configuration space and heterogeneous applications | Take trials with different configurations and decide the next trial direction by Bayesian Optimization model | / |

TABLE 1. Relationships between latest advances and MLN workflow.

## Overview of Recent Advances

Recent breakthroughs of deep learning and other promising machine learning techniques have a non-ignorable influence on new attempts of the network community. Existing efforts have led to several considerable advances in different subfields of networking. To illustrate the relationship between these up-to-date advances and the MLN workflow, in Table 1 we divide literature studies into several application scenarios and show how they perform at each step of the MLN workflow. Without ML techniques, the typical solutions for these advances are involved with time-series analytics [1, 9], statistical methods [1, 5, 7, 8] and rule-based heuristic algorithms [2–5, 10], which are often more interpretable and easier to implement. However, ML-based methods have a stronger ability to provide a fine-grained strategy and can achieve higher prediction accuracy by extracting hidden information from historical data. As a big challenge of ML-based solutions, the feasibility problem is also discussed in this section.

## Information Cognition

Since data are the fundamental resource for MLN, information (data) cognition with high efficiency is critical to capture the network characteristics and monitor network performance. However, due to the complex nature of existing networks and the limitations of measurement tools and architectures, it is still not easy to access some types of data (e.g., trace route and traffic matrix) within acceptable granularity and cost. With its capability for prediction, machine learning can help evaluate network reliability or the probability of a certain network state. As the first example, Internet route measurements help monitor network running states and troubleshoot performance problems. However, due to insufficient usable vantage points (VP) and a limited probing budget, it is impossible to execute each route query because the query may not match any previously measured path or the path may have changed. Sibyl [11] attempts to predict the unseen paths and assign confidence to them by using a supervised machine learning technique called RuleFit.

The learning relies on data acquisition, and MLN also requires a new scheme of data cognition. In MLN, it often needs to maintain an up-to-date global network state and perform real-time responses to client demands, which needs to measure and collect the information in the core network. In order to enable the network to perform diagnostics and make decisions by itself with the help of machine learning or cognitive algorithms, a different network architecture, the Knowledge Plane [12], was presented that can achieve automatic information cognition, which has inspired the following efforts that leverage ML or data-driven methods to enhance network performance.

> Traffic prediction and classification are two of the earliest machine learning applications in the networking field. Because of the well formulated question descriptions and demands from various subfields of networking, studies of the two topics always maintain a certain degree of popularity.

## Traffic Prediction and Classification

Traffic prediction and classification are two of the earliest machine learning applications in the networking field. Because of the well formulated question descriptions and demands from various subfields of networking, studies of the two topics always maintain a certain degree of popularity.

**Traffic Prediction:** As an important research problem, the accurate estimation of traffic volume (e.g., the traffic matrix) is beneficial to congestion control, resource allocation, network routing, and even high-level live streaming applications. There are mainly two directions of research, time series analysis and network tomography, which can be simply distinguished depending on if it conducts traffic prediction with direct observations or not. However, it is expensive to directly measure traffic volume, especially in a large-scale high speed network environment.

Many existing studies focus on reducing the measurement cost by using indirect metrics rather than only trying different ML algorithms. There are two methods to handle this problem. One is to take more human effort to develop sophisticated algorithms by exploring domain-specific knowledge and undiscovered data patterns. As an example, the work in [9] attempts to predict traffic volume according to the dependence between flow counts and flow volume. Another method is inspired by the end-to-end deep learning approach. It takes some easily obtained information (e.g., bits of a header in the first few flow packets) as direct input and extract features automatically with the help of the learning model [10].

**Traffic Classification:** As a fundamental function component in network management and security systems, traffic classification matches network applications and protocols with the corresponding traffic flows. The traditional traffic classification methods include the port-based approach and the payload-based approach. The port-based approach has been proved to be ineffective due to unfixed or reused port assignments, while the payload-based approach suffers from privacy problems caused by deep packet inspection, which can even fail in the presence of encrypted traffic. As a result, machine learning approaches based on statistical features have been extensively studied in recent years, especially in the network security domain. However, it is not easy to consider machine learning as an omnipotent solution and deploy it into a real-world operational environment. For instance, unlike the traditional machine learning application to identify if a figure is a cat or not, it will create a big cost with a misclassification in the context of network security. Generally, these studies range from all-known classification scenarios to a more realistic situation with unknown traffic (e.g., zero-day application traffic [8]). This research roadmap is very similar to the machine learning technology that evolves from supervised learning to unsupervised and semi-supervised learning, which can be treated as a pioneer paradigm to import machine learning into networking fields.

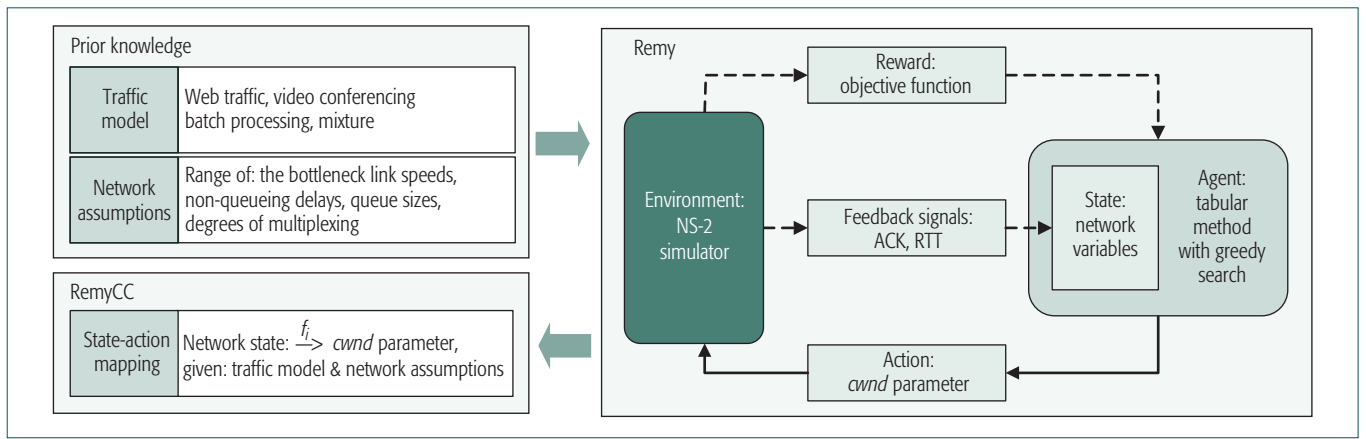

FIGURE 2. Remy's mechanism illustration [3].

### Resource Management and Network Adaption

Efficient resource management and network adaption are the keys to improving network system performance. Some example issues to address are traffic scheduling, routing [2], and TCP congestion control [3, 4]. All these issues can be formulated as a decision-making problem [13]. However, it is challenging to solve these problems with a rule-based heuristic algorithm due to the complexity of diverse system environments, noisy inputs and difficulty in optimizing the tail performance [13]. Specifically, arbitrary parameter assignments based on experiences and action taken following predetermined rules often result in a scheduling algorithm that is understood by people but far from optimal.

Deep learning is a promising solution due to its ability to characterize the inherent relationships between the inputs and outputs of network systems without human involvement. In order to meet the requirements of changing network environments, previous efforts in [2, 14] design a traffic control system with the support of deep learning techniques. Reconsidering backbone router architectures and strategies, it takes the traffic pattern in each router as input and outputs the next nodes in the routing path with Deep Belief Architectures. These advancements unleash the potential of the DL-based strategy in network routing and scheduling. Harnessing the powerful representational ability of deep neural networks, deep reinforcement learning achieves great results in many AI problems.

DeepRM [13] is the first work that applies a deep RL algorithm for cluster resource scheduling. Its performance is comparable to state-of-the-art heuristic algorithms but with less cost. The QoE optimization problem can also benefit from the RL learning methodology. Unlike previous efforts, Pytheas [7] regards this problem as an exploration-exploitation-based problem rather than a prediction-based problem. As a result, Pytheas outperforms state-of-the-art prediction-based systems by lessening the prediction bias and delayed response. From this perspective, machine learning may help achieve the close-loop of "sensing-analysis-decision," especially in wireless sensor networks, where the three actions are separated from each other at present.

Several attempts have been made to optimize the TCP congestion control algorithm using the reinforcement learning approach due to the difficulty of designing a congestion control algorithm that can fit all network states. To make the algorithm self-adaptive, Remy [3] takes the target network assumptions and traffic model as prior knowledge to automatically generate the specific algorithm, which achieves an amazing performance gain in many circumstances. In the offline phase, Remy tries to learn a mapping, i.e., RemyCC, between the network state and the corresponding parameters of the congestion window (cwnd) by interacting with the network simulator. In the online phase, whenever an ACK is received, RemyCC looks up its mapping table and changes its cwnd behavior according to the current network state. The mechanism of Remy is illustrated in Fig. 2. Without the specific network assumptions, a performance-oriented attempt, PCC [4], can benefit from its online-learning nature. Although these TCP-related efforts still focus on decision making, they take the first important step toward automated protocol design.

### Network Performance Prediction and Configuration Extrapolation

Performance prediction can guide decision making. Some example applications are video QoE prediction, CDN location selection, best wireless channel selection, and performance extrapolation under different configurations. Machine learning is a natural approach to predict system states for better decision making.

Typically, there are two general prediction scenarios. First, the system owner has the ability to get various and enough historical data, but it is non-trivial to build a complex prediction model and update it in real time, which requires a new approach exploiting domain-specific knowledge to simplify the problem (e.g., CFA [5] for video QoE optimization). In prior work, CS2P [1] wants to improve video bitrate selection with accurate prediction. It finds that sessions with similar key features may have more related throughput behavior from data analysis. CS2P learns to cluster similar sessions offline and trains different Hidden-Markov Models for each cluster to predict the corresponding throughput given the current session information. CS2P reinforces the correlation of similar sessions in the training process, which outperforms approaches with one single model. This is very similar to the above mentioned traffic prediction problem, since they both passively fit the runtime ground-truth with a certain metric. As another prediction scenario, little historical data exist and it is infeasible to obtain representative data by conducting performance tests due to high trial costs in real network systems. To deal with this dilemma, cherrypick [15] leverages the Bayesian Optimization algorithm to minimize pre-run rounds with a directional guidance to collect representative runtime data of workloads under different configurations.

| Networking application | | Computation speed | | |
| --- | --- | --- | --- | --- |
| Objectives | Specific works | Offline time cost | Online time cost | Device information |
| Network adaption | Ref [2]: routing strategy | Training 100,000 samples with 1000 routers: | When <400 routers: | / |
| | | ~ 100,000 s | >100 ms | Intel i7-6900 K |
| | | ~ 1,000 s | <1 ms | The Nvidia Titan X Pascal |
| | Pytheas [7]: general QoE optimization | Session-grouping: find 200 groups per minute with 8.5 million sessions | Not mentioned | 2.4 GHz, 8 cores and 64 GB RAM |
| | Remy [3]: TCP congestion control | A few hours | Not mentioned | Amazaon EC2 and 80-core and 48-core server |
| Performance prediction | CFA [5]: video QoE optimization | Critical feature learning: ~ 30.1 min every 30–60 min | Quality estimation: ~ 30.7 s every 1–5 min | Two clusters of 32 cores |
| | | | Query response: – 0.66 ms every 1 ms | |
| | CS2P [1]: throughput prediction | Not mentioned | Server side: ~ 150 predictions per second | Intel i7-2.2 GHz, 16 GB RAM, Mac OS X 10.11 |
| | | | Client side: <10 ms per prediction | Intel i7-2.8 GHz, 8 GB RAM, Mac OS X 10.9 |

TABLE 2. Processing time of selective advances.

### Feasibility Discussion

One big challenge faced by ML-based methods is their feasibility. Since many networking applications are delay-sensitive, it is non-trivial to design a real-time system with heavy computation loads. To make it practical, a common solution is to train the model with global information for a long period of time and incrementally update the model with local information in a small time scale [5, 7], which trades off between the computation overhead and information staleness. In the online phase, the common case is to look up the result table or draw the inference with a trained model to make real-time decisions. The processing time in the above advances are selectively listed in Table 2, which shows that ML has practical values with the system well-designed. In addition, the robustness and generalization of a design are also important for feasibility and are discussed later.

From these perspectives, ML in its current state is not suitable for all networking problems. The network problems solved with ML techniques so far are more or less related to prediction, classification and decision-making, while it is difficult to apply machine learning to other types of problems. Other reasons that prevent the application of ML techniques include the lack of labeled data, high system dynamics and high cost brought by learning errors.

### Opportunities for MLN

The prior efforts mostly focus on the generalized concepts of prediction and classification and few can get out of this scope to explore other possible applications. However, with the latest breakthroughs in machine learning and its infrastructures, new potential demands may appear in network disciplines. Some opportunities are introduced as follows.

### Open Datasets for the Networking Community

Collecting a large amount of high quality data that contain both network profiles and performance metrics is one of the most critical issues for MLN. However, acquiring enough labeled data is still expensive and labor intensive even in today's machine learning community. For many reasons, it is not easy for researchers to acquire enough real trace data even if there are many existing open datasets in the networking domain.

This reality drives us to learn from the machine learning community to put much more effort into constructing open datasets like ImageNet. With unified open datasets, performance benchmarks are an inevitable outcome to provide a standard platform for researchers to compare their new algorithms or architectures with state-of-the-art ones. This can reduce the unrepresentative repeated experiments and have a positive effect on academic loyalty. In addition, it has been proved in the machine learning domain that learning with a simulator rather than in a real environment is more effective and with lower cost in RL scenarios [3]. In the networking domain, due

> The current network components are likely to be added based on people's understanding at a time instant rather than a paragon of engineering. There is still enough room for us to improve network performance and efficiency by redesigning the network protocol and architecture.

to the limited accessibility and high test cost of large-scale network systems, simulators with sufficient fidelity, scalability and high running speed are also required. These items contribute to both MLN and further development of the networking domain, and public resources also make it possible for the community to conduct research.

### Automated Network Protocol and Architecture Design

With a deeper understanding of the network, researchers gradually find that the existing network has many limitations. The network system is totally created by human beings. The current network components are likely to be added based on people's understanding at a time instant rather than a paragon of engineering. There is still enough room for us to improve network performance and efficiency by redesigning the network protocol and architecture.

It is still quite difficult to design a protocol or architecture automatically today. However, the machine learning community has made some of the simplest attempts in this direction and has achieved some amazing results, such as letting agents communicate with others to finish a task cooperatively. Other new achievements, e.g., GAN, have also shown that the machine learning model has the ability to generate elements existing in the real world and create strategies people do not discover (e.g., AlphaGo). However, these generated results are still far from the possibility of protocol design. There is great potential and the possibility to create new feasible network components without human involvement, which may refresh human's understanding of network systems and propose some currently unacceptable destructive-reconstruction frameworks.

### Automated Network Resource Scheduling and Decision Making

It is hard to conduct online scheduling with a principle-based heuristic algorithm due to the uncertainty and dynamics of network conditions. In the machine learning community, it has been proved that reinforcement learning has strong capability to deal with decision making problems. The recent breakthrough of Go also proves that ML can make not only coarse but precise decision, which is beyond people's common sense. Although it is not easy to directly apply an exploration-exploitation strategy in highly-varying network environments, reinforcement learning can be a candidate to replace adaptive algorithms of the present network system. Related efforts can refer to [3, 4, 7, 13]. In addition, reinforcement learning is highly suitable for problems where several undetermined parameters need to be assigned adaptively according to the network state. However, these methods introduce new complexity and uncertainty into the network system itself while the stability, reliability and repeatability are always the goals of network design.

Moreover, network scheduling with RL also provides a new opportunity to support flexible objective function and cross-layer optimization. It is very convenient to change the optimization goal just by changing the reward function in the learning model, which is impossible with a traditional heuristic algorithm. Also, the system may perceive high-level application behaviors or QoE metrics as a reward, which may enable adaptive cross-layer optimization without the network model. In practice, it is nontrivial to design an effective reward function. The simplest reward design principle is to set the direct goal that needs to be maximized as the reward. However, it is often difficult to capture the exact optimization objective, and as a result we end up with an imperfect but easily obtained metric instead. In most cases it works well, but sometimes it leads to faulty reward functions that may result in undesired or even dangerous behavior.

### Improving the Comprehension of Network Systems

Network behavior is quite complex due to the end-to-end network design principle, which generates various protocols that have simple actions in the end system but causes nontrivial in-network behavior. From this perspective, it is not easy to figure out what factors can directly affect a certain network metric and can be simplified during an algorithm design process even in a mature network research domain like TCP congestion control. However, with the help of machine learning methods, people can analyze the output of learning algorithms through a posterior approach to find useful insights for us to understand how the network behaves and how to design a high performance algorithm.

For a detailed explanation, DeepRM [13], a resource management framework, is a good example. To understand why DeepRM performs better, the authors find that DeepRM is not work-conserving but decides to reserve room for those yet-to-arrive small jobs, which eventually contributes to reducing job waiting time. For other evidence, refer to CFA [5] and Remy [3] and their following works, which provide insights for key influence factors in video QoE optimization and TCP congestion control, respectively.

### Promoting the Development of Machine Learning

When applying machine learning into networking fields, due to specific requirements of network systems and practical implementation problems, some inherent limitations and other emerging problems of machine learning can be pushed forward to a new understanding stage with the joint efforts of two research communities.

Typically, there are several problems that are expected to be resolved. First, the robustness of machine learning algorithms is a key challenge for applications (e.g., self-driving cars and network operation) in real-world environments where learning errors could lead to high costs. The networking situation often requires hard constraints on the algorithm output and the worst performance guarantee. Second, a model with high generalization ability that can adapt in the high-variance and dynamic traffic circumstances is needed, since it is unacceptable to retrain the model every time the characteristics of network traffic change. Although some of the experiments show that the model trained under a specific network environment can, to some degree, achieve good performance in other environments [3], it is still not easy because most machine learning algorithms assume that the data follow the same distribution, which is not practical in networking environments. In addition, the accountability and interpretability [3] of machine learning algorithms create big obstacles in practical implementations, since many learning models, especially for deep learning, are still black box. People do not know why and how it behaves, hence people cannot interfere with the policy.

## Conclusions

Due to the heterogeneity of networking systems, it is imperative to embrace machine learning techniques in the networking domain for potential breakthroughs. However, it is not easy for networking researchers to take it into practice due to the lack of machine learning related experiences and insufficient directions. In this article, we present a basic workflow to provide researchers with a practical guideline to explore new machine learning paradigms for future networking research. For a deeper comprehension, we summarize the latest advances in machine learning for networking, which covers multiple important network techniques, including measurement, prediction and scheduling. Moreover, numerous issues are still open and we shed light on the opportunities that need further research effort from both the networking and machine learning perspectives.

## Acknowledgment

This work is supported by NSFC (no. 61422206), TNList and the "863" Program of China (no. 2015AA016101). We would also like to thank Keith Winstein from Stanford University for his helpful suggestions to improve this article.

## Biographies

MOWEI WANG received the B.Eng. degree in communication engineering from Beijing University of Posts and Telecommunications, Beijing, China, in 2017. He is currently working toward his Ph.D. degree in the Department of Computer Science and Technology, Tsinghua University, Beijing, China. His research interests are in the areas of data center networks and machine learning.

YONG CUI received the B.E. degree and the Ph.D. degree, both in computer science and engineering, from Tsinghua University, China, in 1999 and 2004, respectively. He is currently a full professor in the Computer Science Department in Tsinghua University. He has published over 100 papers in refereed conferences and journals with several Best Paper Awards. He has co-authored seven Internet standard documents (RFC) for his proposal on IPv6 technologies. His major research interests include mobile cloud computing and network architecture. He served or serves on the editorial boards of IEEE TPDS, IEEE TCC and *IEEE Internet Computing*. He is currently a working group co-chair in IETF.

XIN WANG received the B.S. and M.S. degrees in telecommunications engineering and wireless communications engineering, respectively, from Beijing University of Posts and Telecommunications, Beijing, China, and the Ph.D. degree in electrical and computer engineering from Columbia University, New York, NY. She is currently an associate professor in the Department of Electrical and Computer Engineering, State University of New York at Stony Brook, Stony Brook, NY. Before joining Stony Brook, she was a member of technical staff in the area of mobile and wireless networking at Bell Labs Research, Lucent Technologies, New Jersey, and an assistant professor in the Department of Computer Science and Engineering, State University of New York at Buffalo, Buffalo, NY. Her research interests include algorithm and protocol design in wireless networks and communications, mobile and distributed computing, and networked sensing and detection. She has served on the executive committee and technical committee of numerous conferences and funding review panels, and serves as an associate editor for *IEEE Transactions on Mobile Computing*. She achieved the NSF CAREER Award in 2005 and the ONR Challenge Award in 2010.

SHIHAN XIAO received the B.Eng. degree in electronic and information engineering from Beijing University of Posts and Telecommunications, Beijing, China, in 2012. He is currently working toward his Ph.D. degree in the Department of Computer Science and Technology, Tsinghua University, Beijing, China. His research interests are in the areas of wireless networking and cloud computing.

JUNCHEN JIANG is a Ph.D. candidate in the Computer Science Department at Carnegie Mellon University, Pittsburgh, PA, USA, where he is advised by Prof. Hui Zhang and Prof. Vyas Sekar. He received the Bachelor's degree in computer science and technology from Tsinghua University, Beijing, China, in 2011.